\documentclass[10pt,prl,aps,showpacs,twocolumn,superscriptaddress]{revtex4-2}
\usepackage{graphicx}
\usepackage{color}
\usepackage{dcolumn}
\usepackage{bm}
\usepackage{multirow}
\usepackage{hyperref}

\begin{document}
\preprint{APS/123-QED}
\title{Tissue-Intrinsic Shape Mechanics in Growing Pre-Migratory Tumor Spheroids}
\author{Urban \v Zeleznik}
\affiliation{Faculty of Mathematics and Physics, University of Ljubljana, Jadranska 19, SI-1000 Ljubljana, Slovenia}
\affiliation{Jo\v zef Stefan Institute, Jamova 39, SI-1000 Ljubljana, Slovenia}
\author{Matej Krajnc}
\affiliation{Jo\v zef Stefan Institute, Jamova 39, SI-1000 Ljubljana, Slovenia}
\author{Tanmoy Sarkar}
\email{tsarkar@iisertvm.ac.in}
\affiliation{Jo\v zef Stefan Institute, Jamova 39, SI-1000 Ljubljana, Slovenia}
\affiliation{School of Physics, Indian Institute of Science Education and Research Thiruvananthapuram, Maruthamala PO, Thiruvananthapuram, Kerala 695551, India}
\date{\today}

\begin{abstract}
One of the hallmarks of pre-migratory tumors is the progressive loss of compact morphology. To investigate how tumors may intrinsically regulate their shape during growth, we employ a three-dimensional~(3D) vertex model of multicellular aggregates that incorporates key structural features of tumor spheroids, including its surface, a proliferative rim, and a necrotic core. Focusing exclusively on tumor-intrinsic mechanical interactions, we examine how their collective effects guide morphological evolution {\it en route} to metastasis. We show that spheroids acquire lobulated morphologies through an interplay between differential tensions at the spheroid surface and the living–necrotic interface~(LNI), together with differential growth within the proliferative rim. In addition, spheroid shapes can be substantially modulated by tissue rheological properties emerging from active, cell-scale forces. Our cell- and tissue-scale simulations of tumor morphologies are enabled by a computational framework that overcomes a major limitation of 3D vertex models--the lack of cell-division--by introducing a graph-based polyhedral-division algorithm within the Graph Vertex Model~(GVM).
\end{abstract}

\maketitle

\section{\label{sec:level1}Introduction}
Understanding the physical mechanisms underlying cancer initiation and progression remains a major challenge due to the intrinsic complexity of living tissues~\cite{xin23,massey24,almagro2022tissue}. Uncontrolled cell proliferation gives rise to abnormal cellular aggregates, or tumors, in which cells interact dynamically with one another and with the surrounding extracellular matrix~(ECM)~\cite{quail2013,fiore2020,finger2025}. By regulating organization across multiple spatiotemporal scales, tissue mechanobiology gives rise to pronounced complexity. This includes mechanical heterogeneity, such as spatially patterned stiffness variations in breast and cervical cancers~\cite{liu2021, fuhs2022rigid}, as well as functional differentiation, exemplified by WM793b melanoma spheroids that form distinct proliferative, quiescent, and necrotic layers~\cite{murphy2022,greenspan1972models,stott99}.

In multicellular aggregates, mechanical interactions are often described in terms of effective surface tensions, whose spatial variation plays a central role in shaping the architecture and morphology of both cell monolayers and 3D aggregates~\cite{derganc2009,manning2010,hannezo2014}. Experiments on breast cancer cell lines have demonstrated that the balance between cortical tension and stress fiber–mediated contractility sets the effective surface tension of tumors~\cite{blauth2024different}. In epithelial-like cells, strong cell–cell adhesion lowers this tension by increasing interfacial contact between neighboring cells, whereas free cell surfaces exhibit higher effective tension, favoring compact, rounded spheroids. In contrast, mesenchymal-like cells rely more heavily on stress fiber–mediated contractility and exert larger traction forces on the ECM, resulting in reduced effective surface tension. This reduction promotes surface roughness and facilitates cellular escape from the aggregate~\cite{blauth2024different}. Consistent with these findings, histological analyses of breast cancer and brain tumors indicate that lobulated tumor morphology is strongly associated with advanced disease stage and poorer patient survival outcomes~({\color{blue}Fig.~\ref{Figure1}a} and Ref.~\cite{Yeomans_and_Kas,popadic21}).

Although gene expression and cell signaling regulate tumorigenesis at the molecular level, metastatic initiation is strongly influenced by mechanical tensions at the whole-tumor scale. Computational models of confluent tissues with single-cell resolution are among the modeling frameworks best suited for bridging these scales. In such models, cells are commonly represented as interacting deformable bodies or as continuous fields~\cite{graner1992, camley2014, palmieri2015, mueller2019, alert2020}. Within the vertex model framework, cells are represented as polygons in 2D or polyhedra in 3D~\cite{nagai2001, honda2004, farhadifar2007, fletcher2014, bi2014energy, okuda15, alt2017,lange25,okuda2022,Rastko2017}. Although recent studies have applied vertex models to investigate spheroid mechanics, these efforts have largely been limited either to reduced 2D geometries or to 3D spheroids with fixed cell numbers, only modest deformations, and artificial layers of virtual cells at the spheroid boundary~\cite{zhang2024,parker2025}. 

These restrictions and approximations largely stem from the technical challenges associated with implementing topological transitions—such as cell rearrangements and divisions—in fully 3D settings~\cite{zhang2024,okuda2013,sahu25}. These challenges motivated our recent development of the Graph Vertex Model (GVM), which represents tissue topology as a graph in which nodes correspond to physical entities (vertices, edges, polygonal faces, and cells) and links encode their topological relationships~\cite{sarkar2024}. Within this framework, topological changes in the vertex model are naturally expressed as well-defined graph transformations, which are themselves mathematically represented as graphs, providing an abstract yet intuitive description of local cell-neighborhood reconnections. However, incorporating cell division—an essential process in modeling growing tumor spheroids—has remained an open challenge. In a broader context, the knowledge gaps in the physics of pre-migratory tumor growth extend well beyond the technical limitations of existing vertex models. Indeed, although cancer research spans biology, biochemistry, medicine, physics, and bioengineering, the physical principles governing tumor growth and morphology have historically received far less attention than, e.g., their extensively studied biological counterparts~\cite{nia2020physical}.

Motivated by these gaps, we develop graph transformations encoding cell division within the Graph Vertex Model~(GVM)~\cite{sarkar2024}. This advance enables detailed mechanical simulations of 3D, pre-migratory tumor spheroid growth, explicitly incorporating tumor's surface, an internal necrotic core, and an intermediate viable rim in which cells actively grow and divide. To isolate tissue-intrinsic mechanical contributions from those mediated by ECM, we focus exclusively on intercellular interactions within the spheroid. We systematically vary key physical parameters: (i)~the interfacial tension at the LNI, (ii)~the thickness of the proliferative rim that characterizes heterogeneous tumor growth, (iii)~the spheroid surface tension, and (iv)~the stress-relaxation timescale arising from active tension fluctuations~\cite{curran2017,krajnc2020,yamamoto22}. We investigate how these factors collectively govern tumor morphology, often used as an early indicator of metastatic potential. Our results identify a fundamental tissue-intrinsic triad—surface/interfacial tensions, growth heterogeneity, and active noise—that determines whether growing spheroids remain smooth and compact or instead develop lobulated morphologies characteristic of early invasion.
\section {\label{sec:level4}Results}

\subsection{\label{sec:level8}3D vertex model}

We simulate proliferating cell aggregates using a 3D vertex model, in which each cell is represented as a polyhedron, parameterized by the positions of its vertices,~$\boldsymbol{r}_i$. The topology of the cell–neighbor network is encoded in a GVM graph database, which stores cells along with their vertices, edges, and faces as interconnected nodes~({\color{blue}Fig.~\ref{Figure1}b}, {\color{blue}Methods}, and Ref.~\cite{sarkar2024}).

A cell undergoes shape changes through the displacement of its vertices in response to mechanical forces, as described by a system of dynamical equations. Assuming a friction-dominated environment, the equation of motion for vertex $i$ reads
\begin{equation}
    \label{dyn_eqn}
	\eta\frac{{\rm d}\boldsymbol r_i}{{\rm d}t}=-\nabla_i W+\boldsymbol F_i^{(a)}\>. 
\end{equation}
Here $\eta$ is the friction coefficient, $W$ is the potential energy of the aggregate, and  $\boldsymbol F_i^{(a)}$ denotes the contribution from active forces. 

In our model, we assume surface and bulk contributions to $W$, which reads,
\begin{equation}
    \label{energy}
	W=\sum_{\left<mn\right >}\Gamma_{ mn}A_{mn}+\kappa_V\sum_{m}\left (V_m-V_0\right )^2\>.
\end{equation}
Here $A_{mn}$ and $\Gamma_{mn}$ are the area of the interface between polygonal cells $m$ and $n$ and the effective surface tension on that interface, stemming from cell cortical tension and cell-cell adhesion, respectively~\cite{derganc2009,manning2010,hannezo2014}. The modulus $\kappa_V$ in the second term in $W$ is the cell-incompressibility modulus, which ensures that cell volumes $V_m$ are roughly equal to the preferred volume $V_0$. Adjacent cellular layers with different contractility and adhesion generate finite and distinct interfacial tensions, as measured by laser ablation in premalignant multilayer epithelia~\cite{fiore2020}. By analogy, we assign tension $\Gamma_{mn}=\Gamma_{\rm LNI}$ for polygons at the LNI. Furthermore, the balance between cortical and stress fiber--based contractility within the outermost cells determines the effective surface tension at the tumor’s surface, $\Gamma_{mn}=\Gamma$, for polygons at the surface, where one of the polygons $m$, $n$ is absent~\cite{blauth2024different}. All remaining polygonal faces that do not belong to either the spheroid surface or LNI are assigned an interfacial tension $\Gamma_{mn}=\Gamma_0$.

We choose $V_0^{1/3}$ as the unit of length and $\Gamma_0$ as the unit of tension. Friction and interfacial tension define a timescale $\tau = \eta / \Gamma_0$~[Eq.~(\ref{dyn_eqn})], which is used to nondimensionalize time as $t / \tau \to t$. All surface tensions are rescaled by $\Gamma_0$, such that $\Gamma_{\mathrm{LNI}} / \Gamma_0 \to \Gamma_{\mathrm{LNI}}$ and $\Gamma / \Gamma_0 \to \Gamma$, yielding a dimensionless unit tension at interfaces that are neither at the surface nor at the LNI. Finally, to model highly incompressible cells, we choose $\kappa_VV_0^{4/3}/\Gamma_0=100$.

\subsection{\label{sec:level9}Proliferation profile}
Cell proliferation is coupled to the local availability of a diffusible nutrient~(e.g., glucose) that enters the spheroid from its surface and diffuses inward~\cite{greenspan1972models}. We describe the dynamics of nutrient concentration $c$ by a reaction–diffusion equation, $\partial c/\partial t=D\nabla^2 c - k\,c$, where $D$ and $k$ denote the diffusion coefficient and first-order consumption rate, respectively. Instead of the geometric position within the aggregate, we parameterize the space by the \emph{topological distance} $l$ of a cell from the outer surface, where $l=1$ corresponds to the outermost layer, $l=2$ to the next inward layer, and so on. Discretizing the steady-state equation along $l$ yields a geometric decay of nutrient concentration, $c_l = c_0\,2^{-(l-1)}$, indicating that the available nutrient decreases by one-half with each successive layer, where $c_0$ is the concentration at the aggregate surface~(see {\color{blue}Supplementary Note~1} for details on the derivation). 

In our model, cells are allowed to divide only if their local nutrient concentration exceeds a critical threshold $c_{\mathrm{TH}}$. This nutrient-dependent rule naturally generates a proliferative outer rim and a nutrient-depleted, necrotic core ({\color{blue}Fig.~\ref{Figure1}c} and {\color{blue}Supplementary Note~1}), consistent with experimental observations in tumor spheroids~\cite{murphy2022}. For sufficient nutrients at the outer surface, the division probability per unit time is linearly proportional to the local nutrient level and reads
%
\begin{equation}
    \frac{\Delta p(l)}{\Delta t} = k_0\,2^{-(l-1)}\,\Theta(\lambda - l),
    \label{eq:division_probability_main}
\end{equation}
where $\Theta(\lambda - l)$ is the Heaviside step function and $\lambda$ denotes the proliferative depth (or the thickness of the proliferative rim), defined by $c_{\mathrm{TH}} = c_0\,2^{-(\lambda-1)}$~({\color{blue}Fig.~\ref{Figure1}c}). Because the absolute value of $c_{\mathrm{TH}}$ is not known \emph{a priori} and may vary between cell types and environmental conditions, we systematically vary $\lambda$ to probe how the thickness of the proliferative rim influences the resulting tumor morphology. The parameter $k_0$ is the proliferation rate of surface cells.
In the physically most realistic situation, spheroids grow quasistatically, meaning that $k_0$ is much slower than the timescale of cell-shape relaxation. However, for computational feasibility, we choose a proliferation rate of $k_0=0.01$ which remains sufficiently small to approximate quasistatic growth while keeping the simulations tractable. 

Cells are randomly selected for growth, according to the probability given by Eq.~(\ref{eq:division_probability_main}). Selected cells increase their preferred volume from $V_0$ to $2V_0$ at a constant growth rate $g = 1$ (i.e., the preferred volume doubles in one time unit $\tau$). Upon reaching double the preferred size, the cells divide into two daughter cells along a cleavage plane that is approximately perpendicular to the spheroid surface, but otherwise randomly oriented~({\color{blue}Methods}).

\subsection{\label{sec:levelX}Simulations}

To investigate tumor morphologies, we simulate the growth of spheroids starting from an initially disordered aggregate of $50$ cells~({\color{blue}Methods}). The equations of motion~[Eq.~(\ref{dyn_eqn})] are integrated using a forward finite-difference scheme with a time step of $10^{-3}$. For each combination of parameters, we conduct $50$ independent simulations to obtain well-averaged statistics. Each simulation is initialized with a different random seed to ensure statistical independence and avoid sampling bias. Simulations continue until the simulated spheroids reach a final size of $N_f = 2000$ cells.

\begin{figure*}[htb!]
\includegraphics{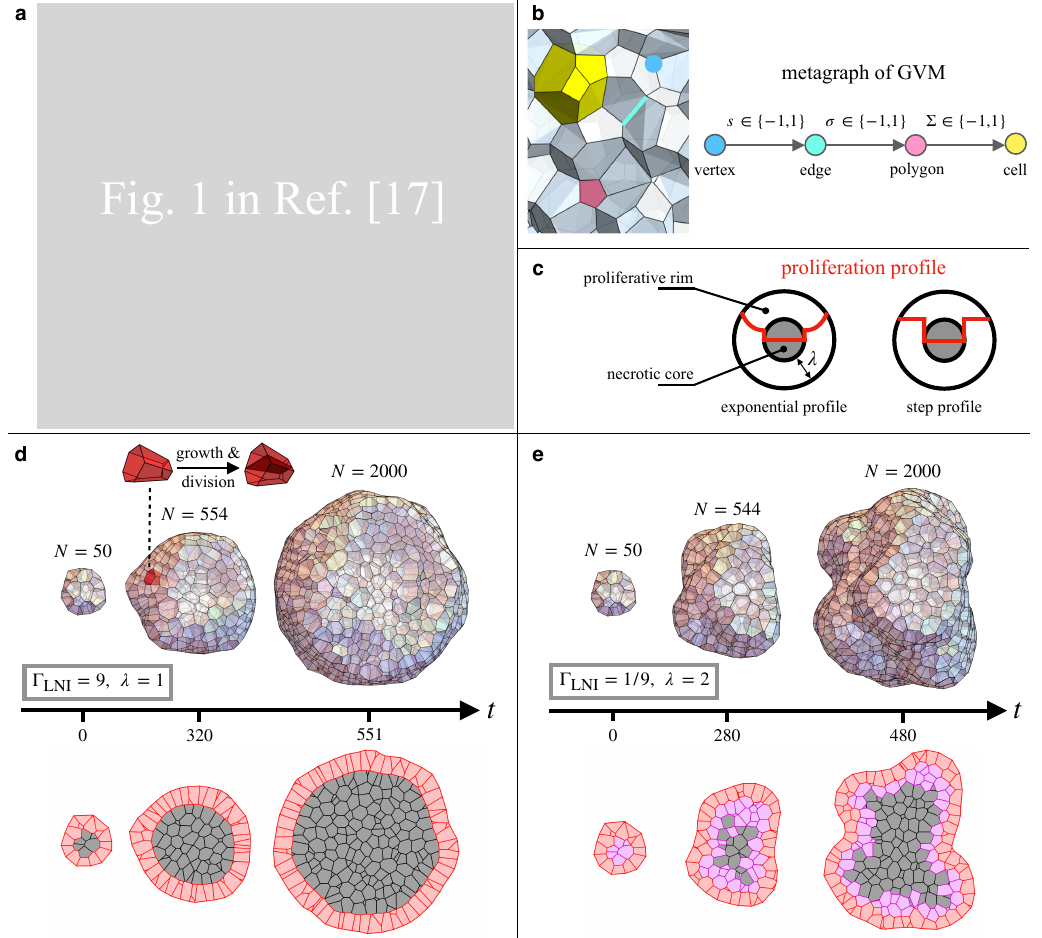}
\caption{\label{Figure1}\textbf{Mechanical model of smooth and lobulated tumor growth.} \textbf{a}~(top)~MR image of a WHO grade I meningioma displaying a regular shape. (bottom)~MR image of a WHO grade III meningioma displaying an irregular shape. Shape is quantified by surface factor, defined as ${\rm SF}=A_{\rm{sphere}}/A$, where $A$ denotes the surface area of the organoid and $A_{\rm sphere}$ the surface area of a sphere with the same volume. \textbf{b}~The metagraph of GVM. On the left, typical vertex, edge, polygon and cell in an aggregate are shaded in blue, cyan, magenta and yellow color, respectively. On the right, the metagraph of GVM is depicted using same color code for nodes which are connected by directed arrows indicating the underlying hierarchy. Contextual properties or signs of vertex $\rightarrow$ edge, edge $\rightarrow$ polygon and polygon $\rightarrow$ cell relationships are indicated by $s$, $\sigma$ and $\Sigma$, respectively. \textbf{c} The two different proliferation profiles (exponential and step distribution) illustrated schematically. \textbf{d,~e} Tumor growth under nutrient-limited conditions and different interfacial tensions between live and necrotic layers. Snapshots show tumor spheroids and their cross-sections, growing from $N = 50$ to $N = 2000$ cells. Inset:~A cell before and after cell division, highlighted in red. The cell grows twice its regular size and divides into two daughter cells. The more spherical spheroid is obtained at $\Gamma=1$, $\Gamma_{\rm LNI} = 9$, and $\lambda = 1$~(panel \textbf{d}), whereas the more lobulated one is obtained at $\Gamma=1$, $\Gamma_{\rm LNI} = 1/9$, and $\lambda= 2$~(panel \textbf{e}). In the cross-sectional view, the dark gray polygons represent necrotic cells and the light red and magenta polygons represent live cells (red polygons represent the first live layer and magenta polygons represent the second live layer).}
\end{figure*}

{\color{blue}Figures~\ref{Figure1}d and e} illustrate representative simulated evolutions of tumor spheroids and their cross-sections during growth. Depending on the interfacial tension at LNI, $\Gamma_{\rm LNI}$, and the thickness of the proliferative rim $\lambda$, spheroids either remain spherical during growth~({\color{blue}Fig.~\ref{Figure1}d}) or develop lobulated morphologies~({\color{blue}Fig.~\ref{Figure1}e}).

\subsection{\label{sec:level6}3D cell-division algorithm}
While simulating individual cell shape changes is relatively straightforward, as it requires only the integration of the equations of motion [Eq.~(\ref{dyn_eqn})], implementing cell-rearrangement events—and cell division in particular—is technically far more challenging. We previously showed that the vertex-model database, encoding the topology of the cellular network, is quite naturally and efficiently handled within a graph-based framework, which we termed the Graph Vertex Model (GVM)~\cite{sarkar2024}. Within this framework, we developed local graph transformations to describe cell rearrangements, however, the implementation of cell division remained an open problem. Here, we generalize GVM by introducing graph transformations that encode the division of a polyhedron into two daughter polyhedra.

\begin{figure*}[htb!]
\includegraphics{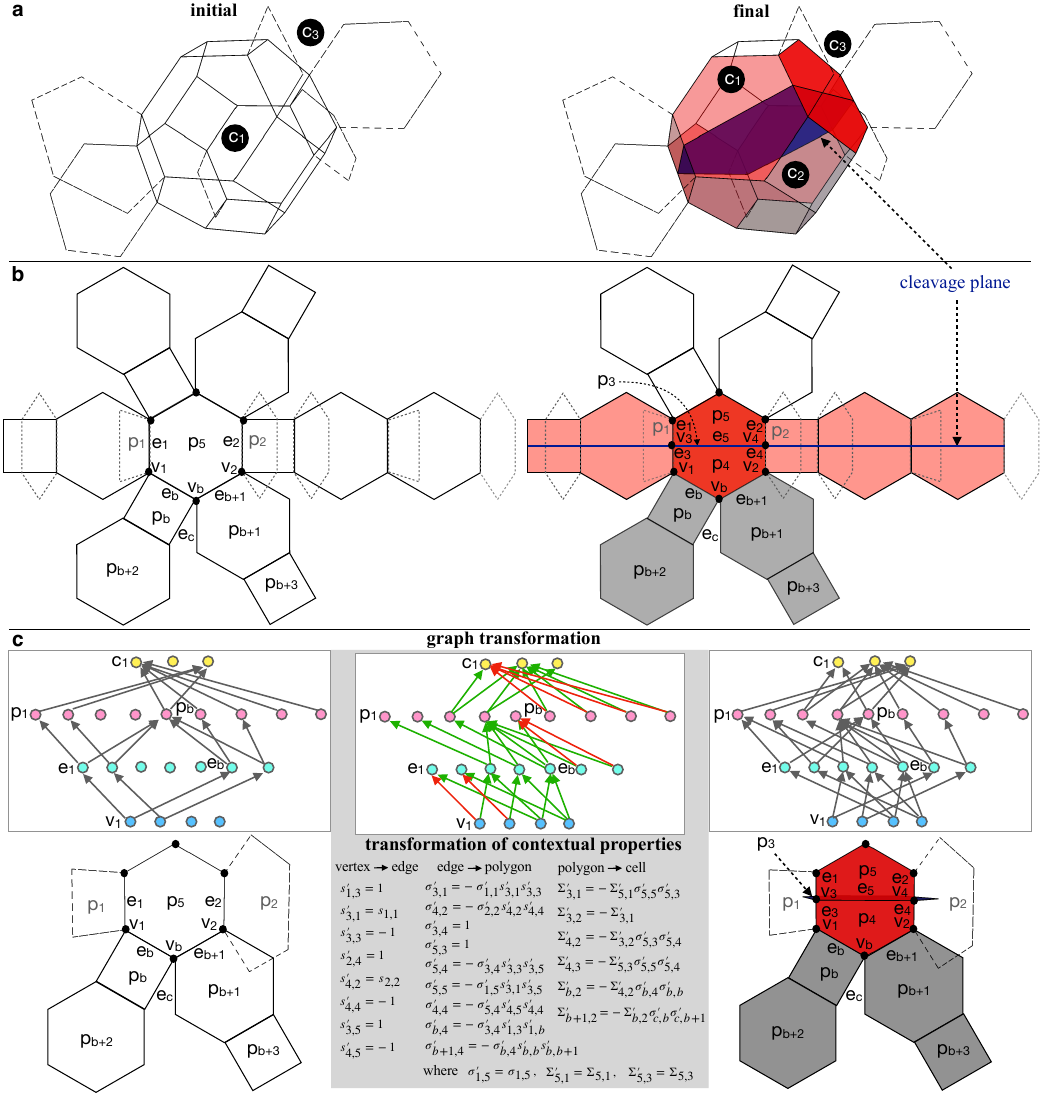}
\caption{\label{fig2}\textbf{Geometry, topology and the graph transformation for a 3D cell division.} \textbf{a} The left and right columns show cells before and after cell division, respectively. Cell $c_1$ divides into two daughter cells, $c_1$ and $c_2$. The cleavage plane is highlighted in dark blue, and new or intersected polygons are marked red. The dark red polygon is shared with cell $c_3$ before division. Polygons above and below the red polygons are white and gray, respectively. \textbf{b}  Polygons of $c_1$ are unwrapped and projected onto 2D before (left) and after (right) division, with the cleavage plane appearing as a straight blue line. Non-$c_1$ polygons are shown with dashed lines. \textbf{c} Subgraphs in the left and right columns highlight the components involved in division from the perspective of the dark red polygon. All nodes follow the same color code as in panel b. The middle column shows the graph transformation for cell division, with deleted and newly created relationships in red and green color, respectively. Transformation of contextual properties of new relationships are compactly shown at the bottom of the middle column. }
\end{figure*}

The algorithm begins by identifying all edges and polygonal faces of the dividing cell that are intersected by the cleavage plane ({\color{blue}Methods} and {\color{blue}Fig.~S1}). For instance, the cleavage plane in {\color{blue}Fig.~\ref{fig2}a} passes through the center of cell $c_1$ dividing it into two daughter cells, $c_1$ and $c_2$. Additionally, the plane intersects edges and polygonal faces marked in red. For clarity, {\color{blue}Fig.~\ref{fig2}b} shows polygonal faces before and after the division, mapped onto a plane. In this 2D projection, the cleavage plane appears as a blue straight line dividing all intersected polygons.

Next, the algorithm arbitrarily chooses one of the intersected polygons, such as the dark red-colored polygon $p_5$ in {\color{blue}Fig.~\ref{fig2}} and performs the following steps iteratively for all other red-colored polygons. First, {\it pattern matching} traverses the graph representing the entire spheroid and isolates a local subgraph containing only the nodes and relationships, representing geometric elements involved in the transformation~({\color{blue}Fig.~\ref{fig2}c}). The division of $p_5$ effects two additional polygons, $p_1$ and $p_2$, which share intersected edges, $e_1$ and $e_2$, respectively, with $p_5$ and belong to the neighboring cell $c_3$. On top of that, polygons of $c_1$, positioned above and below the red patch of polygons~({\color{blue}Fig.~\ref{fig2}}), marked in white and gray, respectively, become part of two different daughter cells. Among these two sets, only one set of polygons, those below the patch ($p_{b+i}$, where $i = 0,\>1,\>2\>...$, etc.), marked in gray, undergoes a change in cell association, as the label $c_1$ of the dividing cell is reassigned to one of the daughter cells. 

The cell $c_1$ is divided by performing a {\it graph transformation} on the initial subgraph generated during the {\it pattern matching}, which deletes and simultaneously creates relationships between nodes as uniquely specified by the graph-transformation graph shown in the middle panel of {\color{blue}Fig.~\ref{fig2}c}. One of the steps of graph transformation is also assigning contextual properties of elements, e.g., headness or tailness of vertices in the context of adjacent edges, $s$, etc., stored in GVM as relationship properties, as described in more detail in {\color{blue}Methods} and Ref.~\cite{sarkar2024}. 

Division of boundary cells is carried out using the exact same graph transformation described above~({\color{blue}Fig.~\ref{fig2}c}, gray panel), but with a smaller matched subgraph and a reduced set of relationship deletions and creations compared to divisions in the bulk. Likewise, when the full 3D transformation is applied to a polygon in a reduced-dimensional 2D vertex model, it is performed as a single-polygon division~({\color{blue}Supplementary Note~2}).

\subsection{\label{sec:level10} Tumor morphologies}

As a result of cell growth and proliferation, tumor spheroids grow and develop diverse, often irregular morphologies. To quantitatively characterize their shape, we use a dimensionless measure of compactness—the reduced volume—defined as the ratio of the spheroid’s volume $V$ to its surface area $A$, 
\begin{equation}
    v = 6\sqrt{\pi}\,\frac{V}{A^{3/2}}\>,
\end{equation}
which equals 1 for a sphere and decreases with increasing surface irregularity~\cite{rozman2020collective}. This definition belongs to a broader class of compactness measures widely used in shape analysis~\cite{bribiesca2008easy}. In two dimensions, a commonly used shape factor is the reduced perimeter, defined as 
\begin{equation}
    S = \frac{P_\mathrm{cs}}{\sqrt{4 \pi A_\mathrm{cs}}}\>,
\end{equation}
where $P_\mathrm{cs}$ and $A_\mathrm{cs}$ denote the perimeter and area of a tumor cross section, respectively~\cite{Yeomans_and_Kas}. In the following, we examine how differential spheroid growth, differential tensions, and tissue rheology jointly determine spheroid morphology, as measured by both $v$ and $S$.

\subsubsection{\label{sec:level15}Interfacial surface tension and rim thickness}
\begin{figure*}[htb!]
\includegraphics{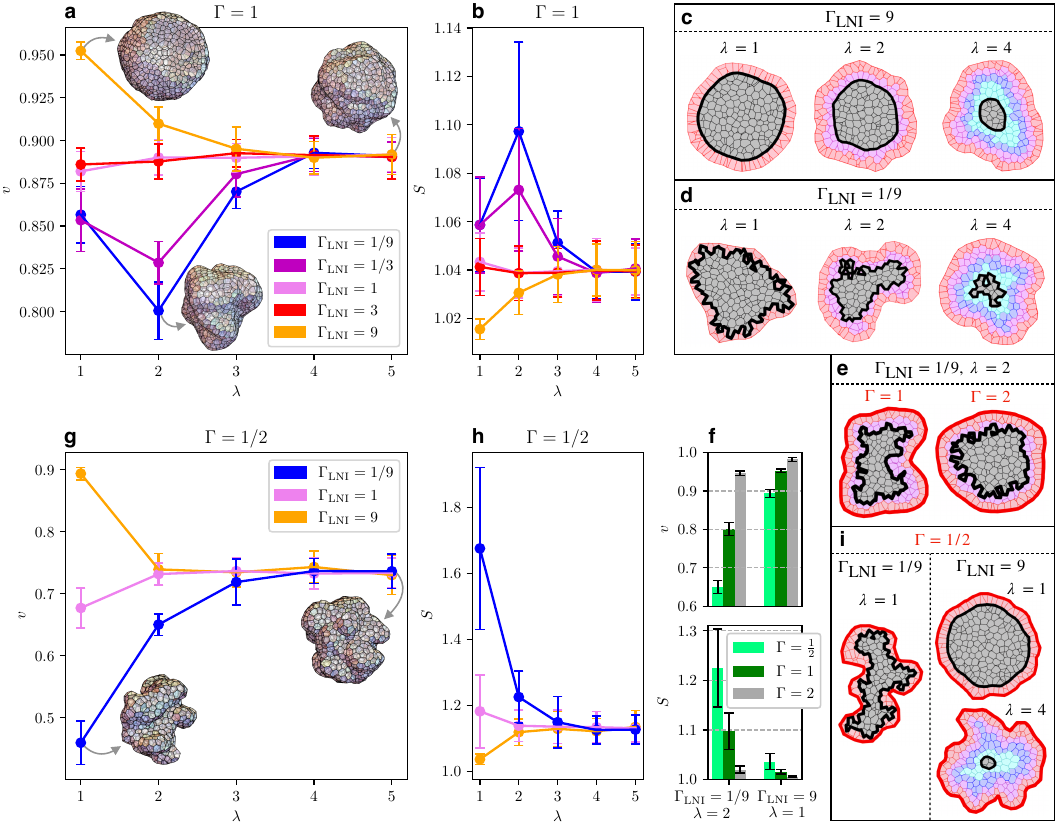}
\caption{\label{Figure3} \textbf{Morphology of simulated tumor spheroids as a function of differential tensions and the thickness of the proliferative rim.} \textbf{a,~b}~The reduced volume $v$ and reduced perimeter $S$ versus the number of live layers $\lambda$ for $\Gamma=1$ and $\Gamma_{\rm LNI} \in \{ 1/9, 1/3, 1, 3, 9 \}$ (blue, purple, pink, red, and orange curves, respectively). The error bars denote the standard deviation across independent simulation instances. \textbf{c-e}~2D spheroid cross-sections corresponding to final states with 2000 cells. The necrotic cells are represented by gray polygons, the outermost live layer of cells is represented by red polygons, the second live layer is represented by magenta polygons, the third by blue, and the fourth by cyan. The cases shown include $\Gamma_{\rm LNI} = 9$ with $\lambda \in \{ 1, 2, 4 \}$~(\textbf{c}), $\Gamma_{\rm LNI} = 1/9$ with $\lambda \in \{ 1, 2, 4 \}$~(\textbf{d}), and $\Gamma_{\rm LNI} = 1/9$ with $\lambda = 2$ and $\Gamma \in \{ 1, 2 \}$~(\textbf{e}). Thick black and red outlines highlight the comparative shape of LNI and spheroid surface as a consequence of $\Gamma_{\rm LNI}$ and $\Gamma$. \textbf{f}~Bar plot of the reduced volume $v$ and the reduced perimeter $S$ for the spheroid surface tensions $\Gamma \in \{0.5, 1, 2\}$ at two different sets of parameters: $\Gamma_{\rm LNI} = 1/9$, $\lambda = 2$ (most irregular tumor case at $\Gamma = 1$) and $\Gamma_{\rm LNI} = 9$, $\lambda = 1$ (least irregular tumor case at $\Gamma = 1$). \textbf{g,~h}~The reduced volume $v$ and reduced perimeter $S$ versus the number of live layers $\lambda$ for $\Gamma=1/2$ and $\Gamma_{\rm LNI} \in \{ 1/9, 1, 9 \}$ (blue, pink, and orange curves, respectively). Insets in panels \textbf{a} and \textbf{g} show representative final simulation snapshots. \textbf{i}~2D spheroid cross-sections corresponding to cases where $\Gamma = 1/2$, shown for $\Gamma_{\rm LNI} = 1/9$ and $\lambda = 1$ as well as $\Gamma_{\rm LNI} = 9$ and $\lambda \in \{ 1, 4 \}$.}
\end{figure*}

We first consider the case in which the spheroid surface tension equals the bulk interfacial tension, i.e., $\Gamma=1$. Our simulations show that tumor morphology is highly sensitive to both the tension at the LNI, $\Gamma_{\rm LNI}$, and the thickness of the proliferative rim, $\lambda$~({\color{blue}Fig.~\ref{Figure3}a,~b}). For large values of $\Gamma_{\rm LNI}$ (orange curves in {\color{blue}Fig.~\ref{Figure3}a,~b}), spheroids adopt compact morphologies characterized by a comparatively high reduced volume $v$ and low reduced perimeter $S$. As $\lambda$ increases, $v$ decreases monotonically while $S$ increases. Due to large $\Gamma_{\rm LNI}$, the LNI acts as a well-defined, smooth interface. When it is separated from the spheroid surface by only a single cell layer, it is mechanically coupled to the surface, thereby suppressing surface undulations~($\lambda=1$-case in {\color{blue}Fig.~\ref{Figure3}c} and {\color{blue}Supplementary Movie~1}). As $\lambda$ increases, the proliferative rim between the surface and the LNI thickens, progressively decoupling the two interfaces~($\lambda=2$- and $\lambda=4$-cases in {\color{blue}Fig.~\ref{Figure3}c} and {\color{blue}Supplementary Movie~2}). As a result, the stabilizing influence of $\Gamma_{\rm LNI}$ weakens and surface undulations become more pronounced~({\color{blue}Fig.~\ref{Figure3}a,~b}). This trend eventually saturates once proliferation extends throughout the spheroid volume and the necrotic core effectively vanishes.

In contrast, at low $\Gamma_{\rm LNI}$~(purple and blue curves in {\color{blue}Fig.~\ref{Figure3}a,~b}), both the reduced volume $v$ and the reduced perimeter $S$—exhibit pronounced non-monotonic dependences on $\lambda$, with extrema corresponding to the most lobulated morphologies occurring at $\lambda=2$. This non-monotonic trend can be explained as follows. Compared to the large-$\Gamma_{\rm LNI}$ regime, the LNI here loses its smooth sheet-like character and instead becomes irregular itself, thereby promoting more strongly distorted spheroid shapes~({\color{blue}Fig.~\ref{Figure3}d}). For $\lambda = 1$ this effect is most pronounced, however, since cell division is restricted to a single outer layer that can expand outward without significant lobulations, the resulting morphology is comparatively smooth~($\lambda=1$-case in {\color{blue}Fig.~\ref{Figure3}d}). When $\lambda=2$, the LNI is still located relatively close to the surface to promote lobulations, but in comparison to the $\lambda=1$-case, here, the differential growth is more pronounced as more cells divide, generating larger compressive tangential stresses, which promote local buckling~($\lambda=2$-case in {\color{blue}Fig.~\ref{Figure3}d} and {\color{blue}Supplementary Movie~3}). With further increasing $\lambda$, the LNI's effect on the surface lobulations becomes more and more limited and spheroids are progressively smoothened~($\lambda=4$-case in {\color{blue}Fig.~\ref{Figure3}d}).

For intermediate $\Gamma_{\rm LNI}$, shapes show only a weak dependence on $\lambda$ and for sufficiently large $\lambda$, all curves converge to universal shape parameters, $v \approx 0.89$ and $S \approx 1.04$~({\color{blue}Fig.~\ref{Figure3}a,~b}).

\subsubsection{\label{sec:levelYY}Spheroid surface tension}
Next, to assess how the surface tension at the outer boundary, $\Gamma$, influences spheroid morphology, we vary it relative to the bulk value. In particular, we consider two representative cases: A reduced tension $\Gamma=1/2$ and an enhanced one $\Gamma=2$. Our simulations reveal that altering $\Gamma$ strongly modifies the shape evolution. 

At large spheroid surface tension, $\Gamma=2$, the surface tends towards a compact spherical shape so as to minimize the surface energy~({\color{blue}Fig.~\ref{Figure3}e}). The most compact spheroids from the $\Gamma=1$ case, obtained at $\Gamma_{\rm LNI}=9$ and $\lambda=1$, become slightly more compact at $\Gamma=2$~({\color{blue}Fig.~\ref{Figure3}f}). Conversely, the lobulated configurations observed at $\Gamma_{\rm LNI}=1/9$ and $\lambda=2$ become substantially more compact, approaching the compactness of the high-tension regime~({\color{blue}Fig.~\ref{Figure3}f}). Apart from being highly regular and compact, the morphologies at high $\Gamma$ become less sensitive to variations in $\Gamma_{\rm LNI}$ and $\lambda$~(gray bars in {\color{blue}Fig.~\ref{Figure3}f}). 

As the surface tension is reduced to $\Gamma=1/2$, spheroid morphology becomes increasingly sensitive to both the living–necrotic interfacial tension $\Gamma_{\rm LNI}$ and the thickness of the proliferative rim $\lambda$~(light green bars in {\color{blue}Fig.~\ref{Figure3}f}). In this regime, the non-monotonic dependence of the reduced volume $v$ and reduced perimeter $S$—observed at $\Gamma=1$ and $\Gamma_{\rm LNI}=1/9$, where the most lobulated shapes occur at $\lambda=2$—is no longer present. Instead, spheroids are most irregular at $\lambda=1$ and small $\Gamma_{\rm LNI}$ ({\color{blue}Supplementary Movie~4}), attaining mean values of $v$ and $S$ as low as $v \approx 0.46$ and as high as $S \approx 1.68$, respectively~({\color{blue}Fig.~\ref{Figure3}g,~h}). Here, the spheroid surface and LNI are both weekly tensed and therefore tend to increase their surface areas~($\Gamma_{\rm LNI}=1/9$-case in {\color{blue}Fig.~\ref{Figure3}i}). This increase of surface areas dominates over the effect of differential growth. As $\lambda$ increases, LNI is progressively buried beneath additional cell layers and its influence diminishes~(blue curves in {\color{blue}Fig.~\ref{Figure3}g,~h}). Nevertheless, the weakly tensed spheroid surface continues to promote surface lobulation across all values of $\lambda$ ({\color{blue}Supplementary Movie~5}), since $\Gamma$ itself is here significantly smaller than the bulk interfacial tension ($\Gamma=1/2<1$). Consequently, while spheroids become more compact with increasing $\lambda$, the reduced volume is relatively low even in the large-$\lambda$ limit, i.e., $v \approx 0.73$, compared to $v\approx 0.89$ for $\Gamma=1$~({\color{blue}Fig.~\ref{Figure3}a,~g}). 

In contrast, $\Gamma=1/2$-spheroids with a single proliferative layer (i.e., $\lambda=1$) are much less lobulated for large interfacial tension at LNI (i.e., $\Gamma_{\rm LNI} = 9$), where LNI becomes the primary determinant of spheroid shape, dominating over the comparatively weak spheroid surface tension $\Gamma$. Again, the influence of LNI diminishes with increasing $\lambda$, which is seen in a rapid convergence of $v$ and $S$ to the universal limit values~({\color{blue}Fig.~\ref{Figure3}g,~h}) and in lobulation that is significantly more pronounced at $\lambda=4$ compared to $\lambda=1$~($\Gamma_{\rm LNI}=9$; $\lambda=1$ and $\lambda=4$-cases in~{\color{blue}Fig.~\ref{Figure3}i}).

\subsubsection{\label{sec:levelXX}Proliferation profile}
Our results demonstrate that, beyond interfacial and surface tensions, the spheroid compactness is importantly determined by spatially heterogeneous growth. To more clearly isolate its effect, we next consider a perturbation, where the differential growth is removed by assuming uniform proliferation profile within the proliferative rim:
\begin{equation}
\frac{\Delta p(l)}{\Delta t} = k_0\,\Theta(\lambda - l)\>,
\label{eq:division_probability_uni}
\end{equation}
such that all cells in the proliferative rim divide with equal probability~({\color{blue}Fig.~\ref{Figure1}c}).

We find that the absence of differential growth within the rim considerably affects the resultant tumor morphology. Firstly, the uniform proliferation suppresses the non-monotonic behavior of $v$ observed at low $\Gamma_{\rm LNI}$ under the decaying profile~(blue curves in {\color{blue}Fig.~\ref{Figure4}a}). Secondly, for intermediate $\Gamma_{\rm LNI}$, where $v$ is nearly insensitive to $\lambda$ in the decaying case, a clear monotonic increase with $\lambda$ emerges under the uniform profile~(pink curves in {\color{blue}Fig.~\ref{Figure4}a}). Thirdly, all curves corresponding to a uniform proliferation profile converge to $v \approx 0.925$ as $\lambda$ increases, whereas the decaying proliferation profile reaches a lower plateau at $v \approx 0.89$~({\color{blue}Fig.~\ref{Figure4}a,~c} and {\color{blue}Supplementary Movie~6}). Overall, the reduced volume $v$ increases substantially across all $\Gamma_{\rm LNI}$ values, except for $\lambda=1$, where the two profiles are equivalent by definition. Altogether, these results indicate that a decaying proliferation profile consistently promotes more irregular, lobulated tumor morphologies than uniform proliferation~({\color{blue}Fig.~\ref{Figure4}b} and {\color{blue}Supplementary Movie~7}). This underscores the central role of differential growth: Higher division rates near the spheroid surface relative to the interior generate an effective area mismatch and the associated stresses, which in turn drive the emergence of undulated shapes characterized by systematically lower reduced volumes $v$.

\subsubsection{\label{sec:levelXXX}Active tension fluctuations}
Stresses generated by differential growth and division may relax on the time scale of our simulations, depending on the rheological properties of the spheroids. These properties can be modulated by a variety of active processes operating at the level of individual cells~\cite{ranft10,bi16,wyatt16,krajnc18}. Such activity gives rise to an effective stress-relaxation time scale, which determines whether the spheroids, over the course of simulation, exhibit solid-like behavior or instead respond as liquid-like materials.

To probe the role of active stress relaxation, we assume active fluctuating force dipoles acting along all edges associated with cells in the proliferating rim, while edges belonging to the necrotic core remain inactive. In 3D cell aggregates, the absence of traction against a substrate precludes the polar self-propulsion commonly invoked in models of active tissues~\cite{bi16,alert2020}, making active force dipoles the most natural choice for introducing an activity-mediated stress-relaxation time scale~\cite{krajnc18,rozman25}. In our model, the contribution of active forces on vertex $i$ reads $\boldsymbol F_i^{(a)}=-\sum_{mn}\gamma_{lmn}(t)\nabla_i L_{lmn}$, where $lmn$ indicates an edge containing the vertex $i$ and also common among cells identified by the indices $l$, $m$, and $n$, with a length of $L_{lmn}$.  The magnitudes of active force dipoles, $\gamma_{lmn}(t)$ fluctuate with time, as prescribed by the Ornstein-Uhlenbeck dynamics:
\begin{equation}
    \label{OU_process}
    \dot\gamma_{lmn}(t)=-\frac{1}{\tau_m}\left (\gamma_{lmn}(t)-\gamma_0\right )+\xi_{lmn}(t)\>.
\end{equation}
Here $\tau_m$ is the relaxation time scale of fluctuations towards a baseline value $\gamma_0$~\cite{curran2017, krajnc2020,yamamoto22}, whereas $\xi_{lmn}(t)$ is a Gaussian white noise with properties $\langle\xi_{lmn}(t)\rangle=0$ and $\langle\xi_{lmn}(t)\xi_{opr}(t')\rangle=\left (2\sigma^2/\tau_m\right )\delta_{lo}\delta_{mp}\delta_{nr}\delta(t-t')$; $\sigma^2$ is the long-time variance of fluctuations and we use $\tau_m=\gamma_0=1$.

\begin{figure*}[htb!]
\includegraphics{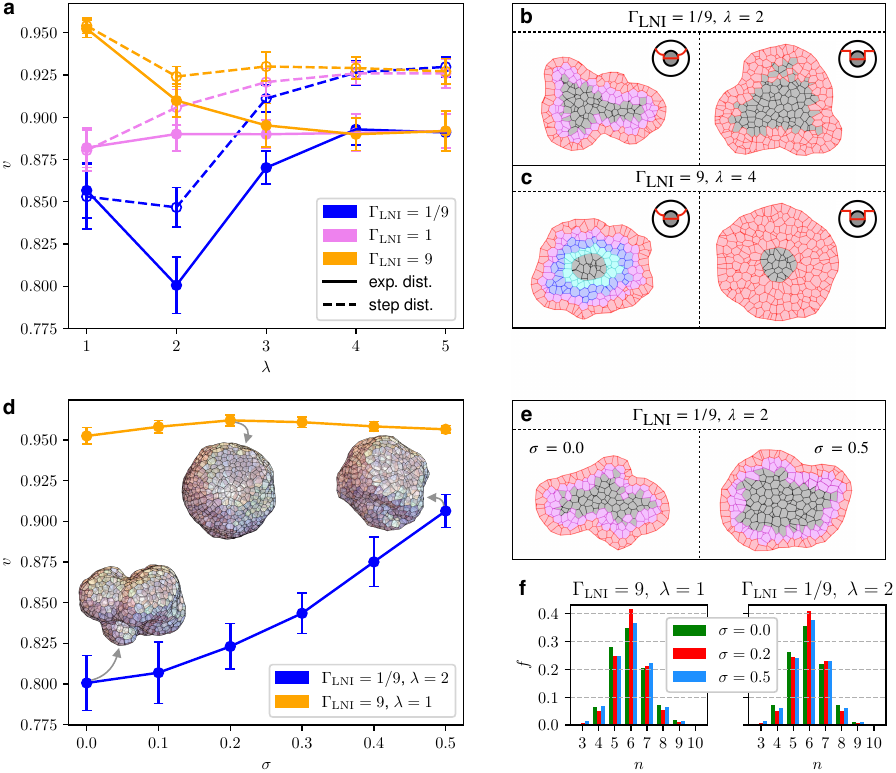}
\caption{\label{Figure4} \textbf{Effect of nutrient distribution profile and amplitude of active fluctuations on tumor morphology.} \textbf{a}~Reduced volume $v$ as a function of the proliferative-rim thickness $\lambda$, for $\Gamma=1$ and $\Gamma_{\rm LNI} \in \{ 1/9, 1, 9 \}$ (blue, pink, and orange curves, respectively). Solid and dashed curves show the results from simulations with exponential and step distibution of nutrients, respectively. The error bars denote the standard deviation across independent simulation instances. \textbf{b,~c}~2D spheroid cross-sections corresponding to final states with 2000 cells at $\Gamma_{\rm LNI}=1/9$, $\lambda=2$ and $\Gamma_{\rm LNI}=9$, $\lambda=4$, comparing the results from exponential and step proliferation profile (left and right, respectively). \textbf{d}~Dependence of the reduced volume $v$ on the amplitude of active tension fluctuations $\sigma$ for two distinct parameter sets with the most extreme tumor shapes previously observed at $\sigma = 0$ and $\Gamma = 1$ ($\Gamma_{\rm LNI} = 1/9$, $\lambda = 2$ and $\Gamma_{\rm LNI} = 9$, $\lambda = 1$). Insets show representative tumor morphologies. \textbf{e}~2D spheroid cross-sections corresponding to final states with 2000 cells at $\Gamma_{\rm LNI}=1/9$, $\lambda=2$ comparing most extreme tension-fluctuations amplitudes, $\sigma=0$ and $\sigma=0.5$ (left and right, respectively). \textbf{f}~Fractions $f$ of surface polygon side numbers $n$ at $\Gamma_{\rm LNI} = 9$, $\lambda = 1$ (left) and $\Gamma_{\rm LNI} = 1/9$, $\lambda = 2$ (right) for $\sigma\in \{ 0, 0.2, 0.5 \}$ (green, red, and blue bars, respectively).  There is an increase of hexagonal surface polygons near the optimal amplitude of edge–force fluctuations $\sigma \approx 0.2$.}
\end{figure*}

For the most lobulated $\Gamma=1$-morphologies observed at $\Gamma_{\rm LNI}=1/9$ and $\lambda=2$, $v$ increases monotonically with the fluctuation strength $\sigma$~(blue curve in {\color{blue}Fig.~\ref{Figure4}d}) and spheroids become markedly more compact compared to the case where active noise is absent. Fluctuations promote shape smoothening because they facilitate cell rearrangements that relax the stresses induced by growth and division~({\color{blue}Fig.~\ref{Figure4}e} and {\color{blue}Supplementary Movie~8}). Nevertheless, even at the largest $\sigma$ considered, $v$ remains well below the values attained for $\Gamma_{\rm LNI}= 9$ and $\lambda=1$, suggesting that surface tensions still have dominating effect in determining the shape.

For high interfacial tension at the LNI~($\Gamma_{\rm LNI} = 9$) and the thinnest proliferative rim ($\lambda=1$), which together yield the most regular, almost perfectly spherical spheroids, the reduced volume $v$ exhibits a weak non-monotonic dependence on the fluctuation strength $\sigma$~(orange curve in {\color{blue}Fig.~\ref{Figure4}d}). Both very low and very high values of $\sigma$ yield slightly less compact morphologies, whereas an intermediate fluctuation amplitude of $\sigma \sim 0.2$ gives rise to the most compact configurations ({\color{blue}Supplementary Movie~9}). This intermediate fluctuation strength, which approximately coincides with the order--disorder transition~\cite{krajnc2020, sarkar2024}, provides an optimal balance: Fluctuations are sufficiently strong to relax growth- and division-induced stresses while maintaining a relatively smooth spheroid surface. The regularization of spheroid's architecture is also seen in relatively ordered organization of polygonal cell faces at the surface~(red bar in {\color{blue}Fig.~\ref{Figure4}f}). At low values of $\sigma$, stress relaxation is too slow to effectively regularize the shape, whereas at very high $\sigma$, spheroids develop rough surfaces due to strongly fluctuating tensions. Consequently, in both regimes the compactness $v$ is slightly reduced compared to the $\sigma \approx 0.2$ case~({\color{blue}Fig.~\ref{Figure4}d}).

\section{\label{sec:level12} Discussion}
We theoretically investigated the evolving morphology of growing pre-migratory tumor spheroids using a 3D vertex model. By integrating a mechanical description of cell–cell interactions based on differential surface tensions with nutrient-limited proliferation and active noise, our framework captures how tissue-intrinsic physical mechanisms determine morphological changes associated with the early loss of shape integrity in pre-migratory tumors. Our results identify a fundamental triad governing spheroid morphology: Interfacial and surface tensions, growth heterogeneity, and tissue rheology.

We found that the interplay between spheroid surface tension, interfacial tension at the LNI, and the thickness of the proliferative rim determines whether growth proceeds via smooth expansion or through the emergence of lobulated shapes, indicative of a transition toward metastatic behavior. Strongly lobulated morphologies arise predominantly when both the surface tension and the LNI interfacial tension are low. Highly lobulated tumors are also observed when surface tension is low but LNI tension is high, provided the proliferative rim is sufficiently thick such that the LNI is buried deep beneath the spheroid surface. By manipulating the proliferation profile, we also probed the role of differential growth. When growth is spatially heterogeneous, stresses accumulate between rapidly dividing outer layers and mechanically distinct inner regions (necrotic core), promoting buckling and shape instability. In contrast, uniform proliferation suppresses these instabilities and yields more compact morphologies, underscoring the central role of differential growth. Highlighting the complexity of underlying mechanics, the interplay between differential growth and LNI tension leads to a non-monotonic relationship between shape compactness and the thickness of the proliferative rim.

We further showed that tissue rheology, encoded here through active tension fluctuations, provides a secondary but essential control. Active fluctuations enable stress relaxation by local cell rearrangements, smoothing irregular shapes. Notably, the non-monotonic response of compactness to activity suggests an optimal fluctuation regime in which stress relaxation is maximized without inducing excessive surface roughness—consistent with active solid–fluid transitions in tissues.

Together, our findings provide a physical interpretation of how growing cell masses may lose their compact shape without invoking biochemical signaling or ECM interactions. Our results demonstrate that variations in growth heterogeneity, mechanical tensions, and stress relaxation are sufficient to reproduce early morphological hallmarks of metastatic behavior. This supports the view that mechanical instabilities are not merely byproducts of malignancy but may actively contribute to its onset~\cite{Punovuori2020}. 

Finally, we generalized GVM by developing the algorithm for cell division. The extended framework establishes a scalable and versatile tool for studying multicellular mechanics in fully 3D dynamically remodeling tissues and overcomes a key limitation of existing implementations of 3D vertex models, i.e., inability to robustly model the dynamics associated with cell division and cell rearrangements. Moreover, the generality of the GVM enables an explicit representation of the aggregate surface, allowing tissue-intrinsic cellular mechanics to be cleanly separated from extracellular effects. Although the present study focuses on very fundamental tissue mechanics, the framework naturally accommodates further extensions including mechanically regulated proliferation and additional forms of active cellular mechanics~\cite{xin23,rozman25}. 

A natural next step is to incorporate explicit ECM mechanics, which will enable systematic investigation of how matrix elasticity, remodeling, and alignment regulate shape stability and facilitate the transition from confined growth to invasive behavior~\cite{pickup2014, mierke2018, cai2024migration, parker2025,zhang2024}. Incorporating these elements will be essential for bridging tumor-scale morphology with cellular-scale mechanotransduction and for strengthening connections between physical modeling, experimental observations, and clinically relevant tumor behavior.

\section{Methods}
\label{method_det}
{\footnotesize

\subsection{\label{subsec:topo} Mapping tissue topology in GVM}
In the vertex model, the potential energy is written as a function of cells' geometric properties, e.g., edge lengths, surface areas of polygons, cell volumes, etc. In order to compute these quantities and their gradients, vertex model needs to keep track of the connections between different geometric entities. In particular, a pair of vertices constitute an oriented edge ${\underline{ \boldsymbol e_j}}$, a series of edges constitutes a polygon ${\underline{ \boldsymbol p_k}}$, and a series of polygons constitutes a cell ${\underline{ \boldsymbol c_l}}$. The main difference in GVM relative to the ``standard'' vertex model is its unique way of storing the topological relations between geometric elements (i.e., vertices, edges, polygons, and cells), described above. In particular, GVM's data model is based on a knowledge graph, whose nodes represent the geometric elements, whereas the links between the nodes represent topological relationships between the elements~({\color{blue}Fig.~\ref{Figure1}b}). This graph data structure and its rules are summarized in the so-called metagraph, which describes in what way different nodes are allowed to be linked to each other. The metagraph of the GVM reads vertex $\rightarrow$ edge $\rightarrow$ polygon $\rightarrow$ cell, where each arrow points from a hierarchically inferior to superior nodes ({\color{blue}Fig.~\ref{Figure1}b}). This hierarchy among different types of nodes keeps all higher-order topological information without creating redundant data, which helps in the straightforward implementation of topological transformations in the cell network, which themselves are represented by graphs~({\color{blue}Fig.~\ref{fig2}c} and Ref.~\cite{sarkar2024}). Since edges are oriented, every edge contains a head- and and a tail vertex. The information about the headness/tailness of vertices is given by a contextual property of vertices, $s\in\left\{-1,+1\right \}$ ($-1$ and $+1$ corresponding to tail and head vertices, respectively). Here, ``contextual'' refers to the fact that headness and tailness of a given vertex are not the absolute properties of that vertex, but rather properties that are bound to a specific edge of which the vertex is part of. In turn, each polygon is assigned a normal vector---the direction associated with the polygon---and its orientation is defined by the right-hand rule. The constituent edges of every polygon are prescribed a contextual property $\sigma\in\left\{-1,+1\right \}$ ($-1$ and $+1$ corresponding to edges pointing in polygon's negative and positive direction, respectively). Similarly, constituent polygons of every cell are prescribed a contextual property $\Sigma\in\left\{-1,+1\right \}$ ($-1$ and $+1$ corresponding to the normal vectors of polygons pointing towards or away from the cell center, respectively). The contextual properties $s$, $\sigma$, and $\Sigma$ are conveniently stored in GVM as relationship attributes~({\color{blue}Fig.~\ref{Figure1}b}).

\subsection{Determining intersected edges and vertices}
To find intersecting edges of the dividing polygon $p_2$, shown in {\color{blue}Fig.~S2}, we inspect each of its edges, for instance, edge $e_1$ connecting $v_1$ and $v_u$ in {\color{blue}Fig.~S1a}. We construct a pair of vectors $\vec{R_1}$ and $\vec{R_2}$ from the center of the polygon $v_{CM}$ to the vertices $v_1$ and $v_u$, respectively. The projections of $\vec{R_1}$ and $\vec{R_2}$ onto the unit normal vector $\hat{n}$ of the cleavage plane---an arbitrary plane containing the axis that connects the center of the cell to the center of the spheroidal aggregate---are of opposite signs. Therefore, $e_1$ is marked as intersected. However, any non-intersecting edge shows projection values with identical signs. Using this criterion, we similarly identify the second intersected edge $e_2$ of the dividing polygon, which is also bisected by the cleavage plane. 

After identifying the intersected edges, $e_1$ and $e_2$, we estimate the locations of intersecting points based on the coordinates of their constituting vertices, center of the polygon and $\hat{n}$. For instance, the location of the intersecting point $v_3$ ({\color{blue}Fig.~S1b}) on the edge $e_1$ reads
\begin{equation}
    \vec{r_1}+(\vec{r_2}-\vec{r_1})\frac{(\vec{r_c}-\vec{r_1})\cdot\hat{n}}{(\vec{r_2}-\vec{r_1})\cdot\hat{n}}\>,
\end{equation}
whereas, the location of $v_4$ on the edge $e_2$ is calculated by replacing $v_1$, $v_u$, and $v_3$ by $v_2$, $v_u'$, and $v_4$, respectively. Here, $v_u'$ is the vertex of $e_2$ other than $v_2$.

\subsection{\label{subsec:ini} Preparation of the initial condition}
We initialize the disordered spheroidal aggregates by sequentially inserting $N = 512$ spheres of diameter $0.5$ using the Voro++ library~\cite{rycroft2009}, producing a dense irregular packing of cells. To impose a free tumor boundary we remove exterior cells by defining a sphere of a small radius ($\approx 3$) at the center of the simulation box and deleting all the cells that have any vertices located outside this sphere, together with their corresponding polygons, edges, and vertices. We finally end up with an approximately spherical aggregate of $N_c = 50$ cells. Additionally, the surface polygons, i.e., those that are not shared by pairs of cells, are transformed such that each cell possess exactly one well-defined exterior face, providing a clean and topologically stable foundation for subsequent simulations of tumor growth and mechanics. All simulations are then performed in a cubic box of side length $L = 100$; spheroids are re-centered to $(L/2, L/2, L/2)$ to guarantee that growth of the tumor remains far from the box boundaries and we thereby avoid spurious boundary effects.

\subsection{Code availability}
\label{code_det}
All simulations and analyses were performed using custom Python code developed by the authors and central to the conclusions of this study. The code was written in Python (3.10 and 3.11) and uses standard scientific libraries including NetworkX, which was used to generate the graph database of topological information. The source code, together with documentation and example configuration files sufficient to reproduce the results, is available at \href{https://github.com/UrbanZeleznik0/Graph-Vertex-Model}{https://github.com/UrbanZeleznik0/Graph-Vertex-Model}. All parameters required to reproduce the datasets are provided in the code repository; no undocumented parameters were used.

}

\bibliography{manuscript}
\section{Acknowledgements}
We acknowledge the financial support from the Slovenian Research and Innovation Agency (research projects J1-3009 and J1-60013, development funding pillar RSF-0106, and research core funding P1-0055). TS acknowledges IISER TVM for support.
\section{Author contributions}
\label{author_contributions}
M.K. designed the research. U.\v Z. performed and analysed numerical simulations under the guidance of T.S. The paper was written by all authors.
\section{Competing interests}
\label{competing_interests}
The authors declare no competing interests.
\newpage

\end{document}


\preprint{APS/123-QED}
\title{Tissue-Intrinsic Shape Mechanics in Growing Pre-Migratory Tumor Spheroids: \\Supplementary Information}
\author{Urban \v Zeleznik}
\affiliation{Faculty of Mathematics and Physics, University of Ljubljana, Jadranska 19, SI-1000 Ljubljana, Slovenia}
\affiliation{Jo\v zef Stefan Institute, Jamova 39, SI-1000 Ljubljana, Slovenia}
\author{Matej Krajnc}
\affiliation{Jo\v zef Stefan Institute, Jamova 39, SI-1000 Ljubljana, Slovenia}
\author{Tanmoy Sarkar}
\email{tsarkar@iisertvm.ac.in}
\affiliation{School of Physics, Indian Institute of Science Education and Research Thiruvananthapuram, Maruthamala PO, Thiruvananthapuram, Kerala 695551, India}
\date{\today}
             
\maketitle

\renewcommand{\thefigure}{S\arabic{figure}}
\setcounter{figure}{0}
\renewcommand{\thesection}{\Alph{section}}
\setcounter{section}{0}
\renewcommand{\thetable}{S\arabic{table}}
\setcounter{table}{0}

\section{Supplementary Figures}
\begin{figure}[htb!]
\includegraphics{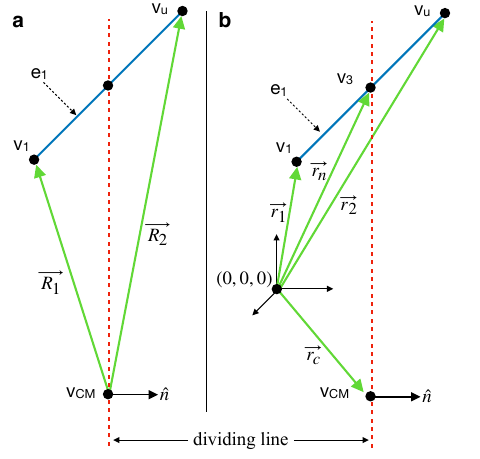}
\caption{\label{S1} \textbf{Identification of an intersecting edge and estimation of the location of a new vertex due to intersection.} The blue solid line represents edge $e_1$, while the red dotted line denotes the dividing plane. $\hat{n}$ indicates the unit vector normal to the dividing plane. $v_1$ and $v_u$ are the vertices of edge $e_1$. \textbf{a}~The green vectors $\vec{R_1}$ and $\vec{R_2}$ extend from the center of the polygon, $v_\mathrm{CM}$, to vertices $v_1$ and $v_u$, respectively. \textbf{b}~Vectors $\vec{r_1}$, $\vec{r_2}$, $\vec{r_n}$, and $\vec{r_c}$ originate from the coordinate system's origin and point to vertices $v_1$, $v_u$, $v_3$, and $v_\mathrm{CM}$, respectively. Vertex $v_3$ is the newly generated vertex resulting from the intersection of the dividing plane with edge $e_1$.}
\end{figure}

\newpage

\begin{figure*}[htb!]
\includegraphics{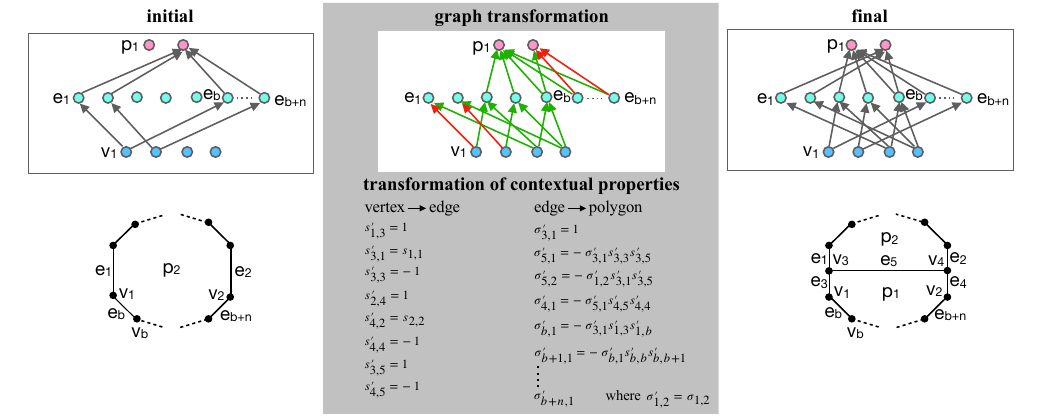}
\caption{\label{S2}\textbf{Graph transformation for a 2D polygon division}. Subgraphs on the left and right show all involved components of a cell before and after cell division, respectively. All nodes follow the same color code as in {\color{blue}Fig.~1b}. The graph in the middle column shows the graph transformation required to perform a polygon division where deleted and newly created relationships are shown in red and green, respectively. Additionally, the contextual properties of these newly created relationships are assigned at the bottom part of the middle column.}
\end{figure*}

\newpage

\begin{figure*}[htb!]
\includegraphics{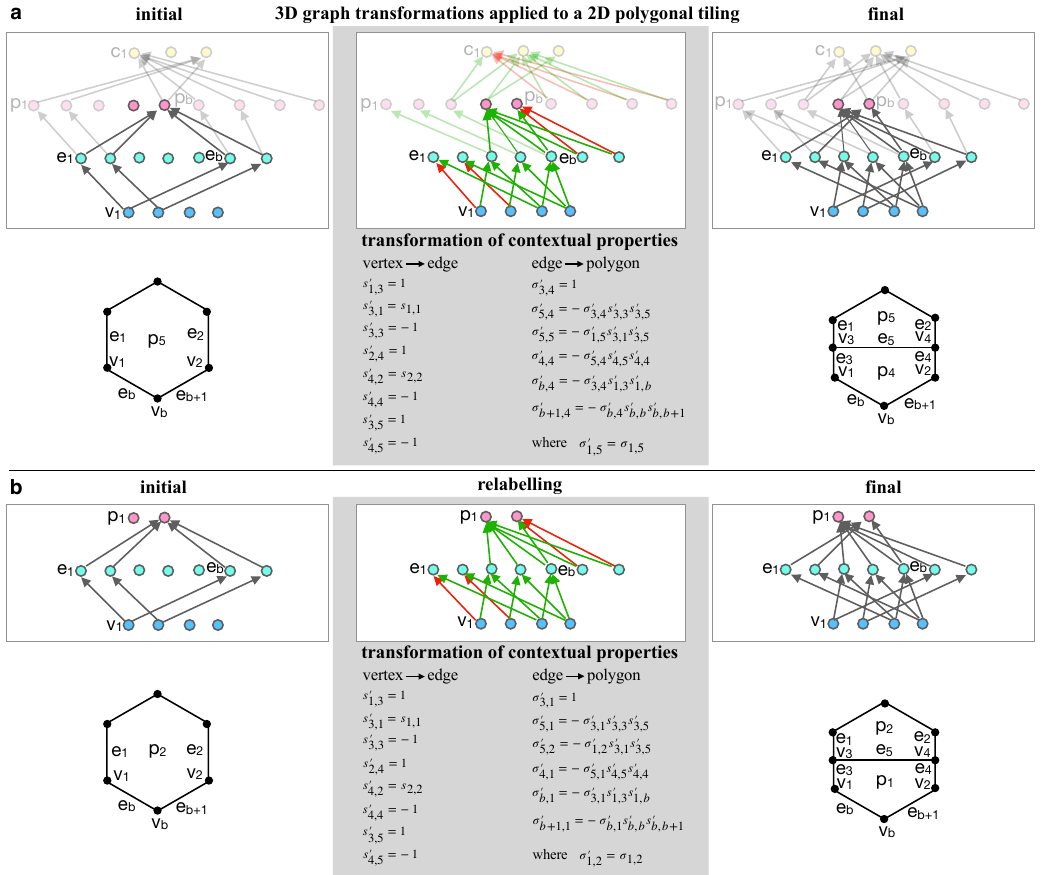}
\caption{\label{S3}\textbf{Graph transformation of polyhedral division in 3D reduces to a 2D polygon division when applied in a planar polygonal tiling}. \textbf{a}~The graph transformation for dividing a polygonal face during 3D cell division~({\color{blue}Fig.~2}) is applied on a 2D planar polygonal tiling. In the initial- (left) and final subgraph (right), and the graph-transformation graph (middle), the nodes and relationships, shown as transparent, indicate unmatched elements due to their absence in the 2D packing. \textbf{b}~After relabelling the polygon nodes, we observe that these three subgraphs exactly match the graphs describing a planar division of a polygon~({\color{blue}Fig.~\ref{S2}}).}
\end{figure*}

\newpage

\section{Supplementary note 1}

\subsection*{Reaction–diffusion description}

We consider that each cell consumes a diffusible nutrient (e.g., glucose) supplied from the spheroid surface~\cite{greenspan1972models}. The nutrient concentration $c$ obeys the standard reaction–diffusion equation,
\begin{equation}
    \frac{\partial c}{\partial t} = D \nabla^2 c - k\,c ,
    \label{eq:diffusion_continuous}
\end{equation}
where $D$ is the diffusion coefficient and $k$ is the first-order consumption rate per unit concentration. In steady state,
\begin{equation}
    \nabla^2 c = \frac{k}{D}\,c ,
    \label{eq:steady_diffusion}
\end{equation}
which expresses the balance between nutrient diffusion and cellular consumption within the tissue.

\subsection*{Topological discretization of layers}

Rather than parameterizing space by geometric radius, we describe the aggregate in terms of the \emph{topological distance} $l = 1,2,3,\ldots$, which counts how many cell layers a given cell lies beneath the spheroid surface.  
The outermost layer corresponds to $l = 1$, the next inward layer to $l = 2$, and so on.  Each layer is separated by a typical cell diameter $\Delta x \sim V_0^{1/3}$, where $V_0$ is the preferred volume of each cell (Eq.~(2) in the main text).  

Discretizing Eq.~(\ref{eq:steady_diffusion}) along this topological coordinate using a second-order central difference yields
\begin{equation}
    \frac{c_{l+1} - 2c_l + c_{l-1}}{(\Delta x)^2} = \frac{k}{D}\,c_l .
    \label{eq:discrete_diffusion}
\end{equation}
Multiplying both sides by $(\Delta x)^2$ and substituting $\Delta x = V_0^{1/3}$ gives
\begin{equation}
    c_{l+1} - 2c_l + c_{l-1} = 
    V_0^{2/3}\frac{k}{D}\,c_l .
    \label{eq:discrete_eq_final}
\end{equation}

\subsection*{Analytic solution of the discrete equation}

We seek a geometric solution of the form $c_l = c_1\,r^{\,l-1}$, representing a constant fractional change in concentration between successive layers,
\begin{equation}
\frac{c_{l+1}}{c_l} = r = \text{const.}
\end{equation}
This assumption is physically reasonable for all interior layers, which experience a similar local balance of diffusion and consumption, except for the outermost layer and the innermost necrotic boundary.

Substituting $c_l = c_1 r^{l-1}$ into Eq.~(\ref{eq:discrete_eq_final}) gives
\begin{equation}
    r^2 - (2 + \alpha)\,r + 1 = 0,
    \qquad \text{with} \quad
    \alpha = V_0^{2/3}\frac{k}{D}.
    \label{eq:recurrence}
\end{equation}
The two roots are
\begin{equation}
    r_{\pm} = \frac{(2 + \alpha) \pm \sqrt{(2 + \alpha)^2 - 4}}{2}.
\end{equation}
The physically admissible solution is the decaying branch $r_- < 1$, ensuring that nutrient concentration decreases toward the spheroid center.

To obtain a compact analytic expression and highlight qualitative behavior, we choose $\alpha = 1/2$, corresponding to a ratio of diffusion to consumption such that
\begin{equation}
    \frac{k}{D} = \frac{1}{2}\,V_0^{-2/3}.
\end{equation}
This choice yields $r_- = 1/2$ and $r_+ = 2$.  
Discarding the unphysical growing mode $r_+$, the nutrient profile becomes
\begin{equation}
    c_l = c_1\,2^{-(l-1)} .
    \label{eq:c_decay}
\end{equation}

The outermost concentration $c_1$ is determined by the boundary condition at the spheroid surface, where the concentration equals that of the external medium, $c_0$.  Substituting $c_1 = c_0$ gives
\begin{equation}
    c_l = c_0\,2^{-(l-1)} ,
    \label{eq:c_decay_updated}
\end{equation}
indicating that the nutrient concentration decreases by one-half for each successive cell layer.

\subsection*{Division threshold and proliferation probability}

Cells are permitted to divide only if their local nutrient concentration exceeds a critical threshold $c_{\mathrm{TH}}$. For viable cells ($c_l \ge c_{\mathrm{TH}}$), the division probability per unit time scales linearly with the local nutrient concentration:
\begin{equation}
    \frac{\Delta p(l)}{\Delta t} \propto c_l .
\end{equation}
Using Eq.~(\ref{eq:c_decay_updated}), this relation can be written as
\begin{equation}
  \frac{\Delta p(l)}{\Delta t} =
  \begin{cases}
    k_0\,2^{-(l-1)}, & c_l \ge c_{\mathrm{TH}},\\[4pt]
    0,                            & c_l < c_{\mathrm{TH}},
  \end{cases}
  \label{eq:division_probability}
\end{equation}
where $k_0$ denotes the \emph{outer-layer division rate} at $l=1$ under the external concentration $c_0$.  

In this work, the absolute value of $c_{\mathrm{TH}}$ is not prescribed {\it a priori} because its magnitude may vary strongly between cell types and their conditions. Instead, we express the threshold in terms of a \emph{topological depth} $\lambda$, defined through
\begin{equation}
    c_{\mathrm{TH}} = c_0\,2^{-(\lambda-1)} .
    \label{eq:threshold_layer}
\end{equation}
This formulation specifies the number of proliferative layers near the surface while maintaining an identical nutrient-dependent division profile across all spheroids:
\begin{equation}
    \frac{\Delta p(l)}{\Delta t} = k_0\,2^{-(l-1)}\,\Theta(\lambda-l),
    \label{eq:complete_eqn}
\end{equation}
where $\Theta(x)$ is the Heaviside function.  
Varying $\lambda$ thus systematically probes how the \emph{extent of the proliferative rim} influences the spheroid morphology, without altering the intrinsic kinetics of cell division.
In our simulations, we set $k_0 = 0.01$ as a nominal surface rate to facilitate comparisons across conditions.  
We also choose a $k_0$ value small enough so that the timescale of spheroid growth is much slower than the timescale of tissue relaxation.

Equations~(\ref{eq:c_decay_updated})–(\ref{eq:complete_eqn}) establish a mechanistic link between nutrient diffusion and spatially heterogeneous cell proliferation. At steady state, diffusion from the spheroid surface cannot fully compensate for local consumption, leading to an exponential (geometric) decay of nutrient concentration with depth. This decay naturally partitions the tissue into distinct regions~\cite{greenspan1972models, freyer1986oxygen}: an outer proliferative rim where $c_l \ge c_{\mathrm{TH}}$, and an inner necrotic core where nutrient levels are insufficient to sustain cell division or viability.

\section{Supplementary Note 2}
\subsection*{Cell division in 2D polygonal tiling}
In a 2D polygonal tiling, each cell is represented by a polygon, i.e., polygon $p_2$ in the left panel of {\color{blue}Fig.~\ref{S2}}. After cell division, polygon $p_2$ splits into two daughter polygons $p_1$ and $p_2$, which either have symmetric or asymmetric sizes, depending on the position and orientation of the division plane. This plane (line in 2D), denoted by $e_5$ in {\color{blue}Fig.~\ref{S2}}, intersects two edges of $p_2$~(i.e., edges $e_1$ and $e_2$ in {\color{blue}Fig.~\ref{S2}}).

The algorithm for dividing a polygon begins by identifying two intersected edges of the dividing polygon $p_2$. Next, the locations of the new vertices, generated due to the intersection of these edges with $e_5$, are calculated. Given these new vertices, each intersected edge, $e_1$ and $e_2$, generates pairs of daughter edges. For instance, the edge $e_1$ generates a pair $e_1$ and $e_3$, whereas, the edge $e_2$ generates a pair $e_2$ and $e_4$~({\color{blue}Fig~\ref{S2}}). After that, topological changes are performed in two consecutive steps as follows.

{\bf(i) Pattern matching} retrieves a small subgraph from the entire graph of GVM. This subgraph contains all the relevant nodes involved in cell division and relationships between them. These data are found by traversing the entire graph database and filtering the results using specific graph-database queries and logical statements as described in more detail in Ref.~\cite{sarkar2024}.

For instance, given the intersected edge $e_1$ and one of its vertices $v_1$, the pattern matching identifies the edge $e_b$ which shares $v_1$ with $e_1$ and also belongs to polygon $p_2$ ({\color{blue}Fig.~\ref{S2}}). Similarly, the edge, next to $e_b$ in the opposite direction of $e_1$ is identified, which also belongs to the polygon $p_2$, eventually collecting all $e_{b+i}$ edges of $p_2$ where $i = 0,\>1,\>2, ..., n$, spanning between $e_1$ and the other intersected edge $e_2$~({\color{blue}Fig.~\ref{S2}}). All the collected nodes and relationships between them construct the initial sub-graph representing the local network topology before the division~(upper left panel of {\color{blue}Fig.~\ref{S2}}). 

{\bf(ii) Graph transformation} converts the initial subgraph to perform the cell division by creating and deleting relationships as shown in the upper-middle panel of {\color{blue}Fig.~\ref{S2}}, where the deleted and newly created relationships are indicated by red and green arrows, respectively. For instance, at the vertex-edge level, relationships from $v_1$ to $e_1$ and $v_2$ to $e_2$ are deleted, whereas two relationships are created from $v_1$ to $e_3$ and from $v_2$ to $e_4$. Furthermore, the vertices $v_3$ and $v_4$, which become connected to the GVM graph through triple newly created relationships, are the intersections points of the division plane with the intersecting edges. The vertex $v_3$ is shared among $e_1$, $e_3$ and $e_5$ while $v_4$ is the common vertex between $e_2$, $e_4$ and $e_5$. At the next hierarchical level, i.e., edge-polygon level, relationships between $e_{b+i}$ and $p_2$ are deleted, but new relationships are established between $e_{b+i}$ edges and one of the newly created polygons $p_1$. Additionally, the two newly generated edges $e_3$ and $e_4$ create two new relationships to $p_1$. Finally, relationships are created between $e_5$ and the two polygons $p_1$ and $p_2$.


Contextual properties (i.e., signs) $s$ and $\sigma$ are assigned to all newly created relationships~(see Ref.~\cite{sarkar2024} and {\color{blue}Methods} in the main text for further details). Some of these properties are calculated from the relationships before division while some can be chosen arbitrarily. The equations in the middle panel of {\color{blue}Fig.~\ref{S2}} summarize how $s$- and $\sigma$-values are calculated; the prime symbol indicates contextual properties after the division. For instance, $s'_{3,1}$, the sign of vertex $v_3$ in the context of edge $e_1$ is assigned the same sign as that of $v_1$ in the context of the edge $e_1$ before the division, since vertex $v_3$ merely replaces the role of $v_1$ in the context of $e_1$. In contrast, $s'_{3,5} = 1$ is arbitrarily chosen, since there is no restriction on vertices of the dividing edge $e_5$. Among contextual properties that need to be calculated, the sign of $e_5$ in the context of $p_2$, $\sigma'_{5,2}$ depends on three signs: (1) $\sigma'_{1,2}$, the sign of $e_1$ in the context of $p_2$, (2) $s'_{3,1}$, the sign of $v_3$ in the context of $e_1$ and, (3) $s'_{3,5}$, the sign of $v_3$ in the context of $e_5$. In short,  $\sigma'_{5,2}$ receives the same or opposite sign of $\sigma'_{1,2}$ depending on the dissimilarity or similarity of $s'_{3,1}$ and $s'_{3,5}$, respectively.

\subsection*{Generalization of cell division in 2D and 3D} 

During cell division in three dimensions, polygons intersected by the division plane are split into pairs of daughter polygons, analogous to division events in two dimensions. This observation suggests that, as in cell rearrangements~\cite{sarkar2024}, the same graph transformation used to describe cell division in 3D may also be applied to a 2D polygonal tiling to model polygon division. To demonstrate this equivalence, we apply the 3D division algorithm to a 2D polygonal tiling, as illustrated in {\color{blue}Fig.~\ref{S3}}. Due to the absence of cells and a few accompanying polygons, only a part of the 3D graph-transformation graph is matched ({\color{blue}Fig.~\ref{S3}a}). The unmatched nodes and relationships are shown in transparent shades. After relabelling $p_5$ and $p_4$ polygons in that subgraph by $p_2$ and $p_1$, respectively, the resulting graph corresponds to the same transformation graph as in the case of polygon division in 2D polygonal tiling~({\color{blue}Fig.~\ref{S3}b and \ref{S2}}), thereby confirming that the graph-transformation graph that performs a cell division in 2D is a subgraph of the more general 3D transformation graph.

\section*{Supplementary Movies}
The supplementary movies illustrating tumor spheroid growth and morphological evolution are available at \href{https://github.com/UrbanZeleznik0/Graph-Vertex-Model}{https://github.com/UrbanZeleznik0/Graph-Vertex-Model}.

\vspace{0.5em}

\noindent \textbf{Supplementary Movie 1.}
Simulation of a growing tumor spheroid at $\sigma=0$, $\Gamma=1$, $\Gamma_{\mathrm{LNI}}=9$, and $\lambda=1$, shown for the whole three-dimensional tumor (left) and its cross section (right).

\vspace{0.3em}
\noindent \textbf{Supplementary Movie 2.}
Simulation of a growing tumor spheroid at $\sigma=0$, $\Gamma=1$, $\Gamma_{\mathrm{LNI}}=9$, and $\lambda=5$ with the exponential proliferation profile, shown for the whole three-dimensional tumor (left) and its cross section (right).

\vspace{0.3em}
\noindent \textbf{Supplementary Movie 3.}
Simulation of a growing tumor spheroid at $\sigma=0$, $\Gamma=1$, $\Gamma_{\mathrm{LNI}}=\tfrac{1}{9}$, and $\lambda=2$ with the exponential proliferation profile, shown for the whole three-dimensional tumor (left) and its cross section (right).

\vspace{0.3em}
\noindent \textbf{Supplementary Movie 4.}
Simulation of a growing tumor spheroid at $\sigma=0$, $\Gamma=\tfrac{1}{2}$, $\Gamma_{\mathrm{LNI}}=\tfrac{1}{9}$, and $\lambda=1$, shown for the whole three-dimensional tumor (left) and its cross section (right).

\vspace{0.3em}
\noindent \textbf{Supplementary Movie 5.}
Simulation of a growing tumor spheroid at $\sigma=0$, $\Gamma=\tfrac{1}{2}$, $\Gamma_{\mathrm{LNI}}=\tfrac{1}{9}$, and $\lambda=5$ with the exponential proliferation profile, shown for the whole three-dimensional tumor (left) and its cross section (right).

\vspace{0.3em}
\noindent \textbf{Supplementary Movie 6.}
Simulation of a growing tumor spheroid at $\sigma=0$, $\Gamma=1$, $\Gamma_{\mathrm{LNI}}=9$, and $\lambda=4$ with the step proliferation profile, shown for the whole three-dimensional tumor (left) and its cross section (right).

\vspace{0.3em}
\noindent \textbf{Supplementary Movie 7.}
Simulation of a growing tumor spheroid at $\sigma=0$, $\Gamma=1$, $\Gamma_{\mathrm{LNI}}=\tfrac{1}{9}$, and $\lambda=2$ with the step proliferation profile, shown for the whole three-dimensional tumor (left) and its cross section (right).

\vspace{0.3em}
\noindent \textbf{Supplementary Movie 8.}
Simulation of a growing tumor spheroid at $\sigma=0.5$, $\Gamma=1$, $\Gamma_{\mathrm{LNI}}=\tfrac{1}{9}$, and $\lambda=2$ with the exponential proliferation profile, shown for the whole three-dimensional tumor (left) and its cross section (right).

\vspace{0.3em}
\noindent \textbf{Supplementary Movie 9.}
Simulation of a growing tumor spheroid at $\sigma=0.2$, $\Gamma=1$, $\Gamma_{\mathrm{LNI}}=9$, and $\lambda=1$, shown for the whole three-dimensional tumor (left) and its cross section (right).

\bibliography{manuscript}